# Data-Driven Linearized AC Power Flow Model with Regression Analysis

Xingpeng Li, *Student Member*, *IEEE* and Kory Hedman, *Member, IEEE*

*Abstract*— **Full AC power flow model is an accurate mathematical model for representing the physical power systems. In practice, however, the utilization of this model is limited due to the computational complexity associated with its non-linear and non-convexity characteristics. An alternative linearized DC power flow model is widely used in today's power system operation and planning tools. However, when reactive power and voltage magnitude are of concern, DC power flow model will be useless. Therefore, a linearized AC (LAC) power flow model is needed to address this issue. This paper first introduces a regular LAC model; subsequently, with the advance in regression analysis technique, a data-driven linearized AC (DLAC) model is proposed to improve the regular LAC model. Numerical simulations conducted on the Tennessee Valley Authority (TVA) system demonstrate the performance and effectiveness of the proposed DLAC model.**

## I. INTRODUCTION

POWER system is one of the most complicated physical networks in the world. Almost all electricity demands are served via power systems. It is of utmost importance to study the fundamentals of power systems. Power flow analysis is one of the essential fundamentals in the power system domain.

Power flow model is involved in a great number of problems and applications in power systems. For instance, it is incorporated into the formulations of state estimation, security-constrained economic dispatch, security-constrained unit commitment, transmission maintenance scheduling, and transmission expansion planning. Hence, power flow studies are remarkably important for power system planning as well as power system operations. The two most popular power flow models are the full AC model and the linearized DC model.

The full AC power flow model, following Kirchhoff's circuit laws and Ohm's law, can accurately represent the associated physical system. This model captures all electric variables of interest, including active power, reactive power, phase angle and voltage magnitude. However, it is not uncommon to observe the divergence of AC power flow problems even with commercial software such as PSS/E and DSATools.

Though a number of algorithms were proposed in literature to improve the convergence performance of the AC model [1]-[2], it still remains to be an unresolved issue. The computational complexity due to its non-linearity and non-convexity makes it impossible to use the AC model in solving a variety of optimization problems. For instance, though the classical formulation for AC economic dispatch was first created by Carpentier in 1962 [3], no robust and reliable algorithm has been developed since then to solve the problem in a timely manner due to its non-linearity, non-convexity and large-scale features. As a result, the industry still uses a linearized DC power flow model today.

To relieve computational burden, the simplified traditional DC power flow model is adopted when only active power and phase angle are of concern. The DC model is simple, efficient and reliable; thus, it is widely used in the power industry and many power system applications [4]-[6]. For instance, instead of the AC model, the DC model is employed in the day-ahead energy markets and real-time energy markets.

The DC model is a good approximation of the AC model in terms of active power solution for high voltage transmission networks, of which the X/R ratios are typically very high. However, DC power flow may fail to perform properly in some scenarios. For instance, the work in [7] shows that DC model based cascading failure simulators fail to capture the power system behaviors in several circumstances. Though the average error for active power is limited to 5%, significant errors are still observed on several individual lines [8]. Three DC power flow models are investigated in [9]. It is concluded that the α-matching model has the most accurate results. However, α-matching method is hot-start and thus, its utilization is only restricted to near-real-time applications with the knowledge of system initial status. Furthermore, DC model cannot be utilized in the cases that voltage magnitude and reactive power are of interest.

Full AC power flow model is accurate, but its computational complexity and unstable characteristics restrict its utilization. DC power flow model can reduce the computational burden significantly, but it may suffer inaccuracy issue and report no information regarding reactive power and voltage magnitude. Therefore, for situations when the reactive power and voltage information are needed while the solution time is limited, a linearized AC (LAC) power flow model that can capture reactive power and voltage magnitude is desired.

Three linear-programming AC power flow models, derived with polyhedral relaxation and Taylor series, are proposed in [10]; numerical simulation demonstrates the effectiveness of the proposed models. Another linear approximation of the AC

Xingpeng Li and Kory Hedman are with School of Electrical, Computer and Energy Engineering, Arizona State University, Tempe, 85281, AZ, USA (email: xingpeng.li@asu.edu; kwh@myuw.net).

model is presented in [11]. The error of this approximation is less than 6% for voltage magnitude. It is worth noting that the approximation error of the proposed model in this paper is only about 1%. A linear relaxation of AC power flow model using polynomial optimization is proposed in [12]. The linearized model is applied to transmission planning problem, and the case studies demonstrate its capability of obtaining approximate solutions in a reasonable time. However, the solution is not checked and compared with the accurate full AC model. An iterative linear power flow method proposed in [13] seems to be accurate and fast; however, this iterative method may fail to converge. All above work [10]-[13] is demonstrated on small-scale standard test cases only, and further efforts are needed to investigate the model accuracy on large-scale practical power systems.

A data-driven linearized AC (DLAC) power flow model is proposed in this paper. This model captures all system state variables including active power, phase angle, reactive power and voltage magnitude. Initially, a regular linearization of AC power flow model is conducted by ignoring higher order terms. Then, coefficients are assigned to the terms that are left in the model. Regression analysis technique is performed to determine those coefficients, which will reflect the system most recent status in the regression model. The philosophy behind this idea is that the system condition does not change significantly in a short timeframe. For instance, the generator voltage setting points typically do not change or change within very narrow ranges, which indicates that voltage magnitude for other neighboring buses would also not change significantly at least in a short timeframe. Another noticeable fact is that the voltage magnitudes for high voltage transmission networks are typically higher than one per unit.

The rest of this paper is organized as follows. Section II presents an overview on the power flow model. Section III introduces the regression analysis technique. Section IV discusses the regular LAC model and the proposed DLAC model. Case studies are presented in Section V. Finally, Section VI concludes the paper.

## II. AC Power Flow Model

The power flow problem is the basis of power system analysis. The per unit system and single-line diagram are usually used for simplification. In the power flow studies, the following assumptions are typically made.
- The system is three-phase balanced and thus, only the positive sequence network is of concern.
- The Pi-equivalent circuit model can accurately represent the transmission network.
- The individual generation and load are known except for the generation at the slack bus.

Given these assumptions, the following state variables can be obtained by solving an AC power flow problem through computer programs:
- voltage magnitude and phase angle at each bus,
- active power and reactive power generations at each bus,
- active power flow and reactive power flow in both directions on each branch, and losses on each branch.

Fig. 1 shows the single-line diagram of a typical two-terminal circuit. A power system network consists of a number of those 2-terminal circuits. Note that $P$ denotes active power while $Q$ stands for reactive power. Normally, the power flowing out of one end-bus does not equal to the power flowing into the other bus because of 1) the reactive power produced by transmission lines, and 2) the losses on the branch connected to the two end buses, which means that $P_{ij} \neq -P_{ji}$ and $Q_{ij} \neq -Q_{ji}$.

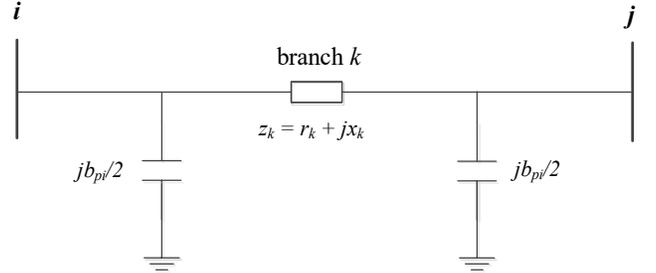

Fig. 1. Single-line diagram of a 2-terminal circuit

The power flow equations for a branch are given below.
$$P_{ij} = V_i^2 g_{ij} - V_i V_j (g_{ij} cos\theta_{ij} + b_{ij} sin\theta_{ij}) \quad (1)$$
$$Q_{ij} = -V_i^2 \left(b_{ij} + \frac{b_{pi}}{2}\right) + V_i V_j (b_{ij} cos\theta_{ij} - g_{ij} sin\theta_{ij}) \quad (2)$$

where, $P_{ij}$ and $Q_{ij}$ denote the active power and reactive power on branch $k$ flowing from bus $i$ to bus $j$ respectively; $\theta_{ij}$ denotes the phase angle difference across this branch; $V_i$ and $V_j$ are the voltage magnitude of bus $i$ and bus $j$ respectively; $b_{pi}$ denotes the susceptance of Pi-equivalent circuit. $g_{ij}$ and $b_{ij}$ are derived from the following equation.
$$y_{ij} = g_{ij} + jb_{ij} = \frac{1}{z_{ij}} = \frac{r_{ij} - jx_{ij}}{r_{ij}^2 + x_{ij}^2} \quad (3)$$

Other important equations for the power flow studies are the nodal power balance constraints as shown in (4) and (5).
$$P_{ig} - P_{id} - \sum_{h \in S_{i+}} P_{ih} + \sum_{h \in S_{i-}} P_{hi} = 0 \quad (4)$$
$$Q_{ig} - Q_{id} - \sum_{h \in S_{i+}} Q_{ih} + \sum_{h \in S_{i-}} Q_{hi} = 0 \quad (5)$$

where $S_{i+}$ denotes the set of buses that are directly connected to bus $i$ when bus $i$ is designated as the sending end, and $S_{i-}$ denotes the set of buses that are directly connected to bus $i$ when bus $i$ is considered as the receiving end; $P_{ig}$ and $Q_{ig}$ represent the total active power and total reactive power produced by the generators at bus $i$, respectively; $P_{id}$ and $Q_{id}$ are the total active power load and total reactive power load at bus $i$ respectively; $P_{ih}$ and $Q_{ih}$ denote the branch active power and reactive power flowing from bus $i$ to bus $h$ respectively.

## III. Regression Analysis

Regression analysis is a widely used statistical technique for estimating the relationships among variables and determining the model for those variables. It typically involves a dependent variable, which is also referred to as a response variable, and one or multiple independent variables that are often called as regressors or predictors. The most popular method is the least squares approximation. A general multiple linear regression model with $k$ regressors is defined as follows:

$$y = \beta_0 + \sum_{j=1}^{k} \beta_j x_j + \varepsilon \qquad (6)$$

where $x_j$ is the $j^{th}$ regressor; $\beta$ denotes the coefficient; and $\varepsilon$ denotes the error.

Provided a sample set of $n$ observations, all parameters $\beta_j$ can be determined as $\hat{\beta}_j$ through regression analysis. The estimated value $\hat{y}_i$ and residual $r_i$ for the $i^{th}$ observation in the sample space are defined as (7) and (8) respectively.

$$\hat{y}_i = \hat{\beta}_0 + \sum_{j=1}^{k} \hat{\beta}_j x_{ij} \qquad (7)$$
$$e_i = y_i - \hat{y}_i \qquad (8)$$

where $y_i$ is the observed value of the $i^{th}$ observation. Residual $e$ denotes the difference between the observed value and the estimated or fitted value while error $\varepsilon$ represents the discrepancy between the observed value and the true value.

After creating a regression model, it is important to (i) conduct model adequacy checking to ensure the model fits the data and (ii) perform model validation to demonstrate the effectiveness of the regression model.

*A. Model Adequacy Checking*

One statistical metric to evaluate the overall model adequacy is the coefficient of determination, which is a percentage number denoted as $R^2$. $R^2$ quantifies how good the regression model is and how much variation can be explained with the regression model. In other words, $R^2$ is the proportion of variation in the response variable explained and predicted by the regressors. 100% indicates that the regression model explains all the variability around the mean. $R^2$ is defined in the equation below,

$$R^2 = 1 - \frac{SS_{res}}{SS_{total}} \qquad (9)$$

where $SS_{res}$ denotes the residual sum of squares and $SS_{total}$ denotes the total sum of squares.

Residual analysis can effectively discover several types of model inadequacies and measure how good the regression model fits the data [14]. Scaled residuals such as standardized residuals may be a better analysis technique to find outliers and analyze the regression model. The standardized residuals with zero mean and approximately unit variance is defined in the equation below,

$$d_i = e_i/\sqrt{MS_{Res}} \qquad (10)$$

where $MS_{Res}$ is the residual mean square that estimates the average variance of residuals.

Another commonly used scaled residuals are R-student residuals defined in (11) below. R-student residuals are often used since their variance is constant.

$$t_i = e_i/\sqrt{S_{(i)}^2(1 - h_{ii})} \qquad (11)$$

where $h_{ii}$ is an element of the hat matrix and $S_{(i)}^2$ denotes the estimate of variance with the $i^{th}$ observation being moved from the dataset.

A plot of the residuals against the fitted values can help detect several types of model inadequacy. If the residuals in the plot are contained in a horizontal band, then, there is no indication of model deficiency. If the residuals against fitted values form a pattern such as funnel, double bow or nonlinear, then, it indicates defects may exist in the regression model and further investigation is required.

*B. Model Validation*

In the regression analysis domain, the best fitted model to the sample dataset may not accurately describe the variables relationship for a different scenario. One key concern is the danger of extrapolation. Though a regression model is often used for extrapolation, it does not apply to the power flow analysis since the power system status does not change significantly especially in a short time frame. The case studies section in this paper demonstrate that the proposed regression model for linearized power flow equations has very similar performance in different system scenarios.

Multicollinearity may occur when regressors are highly linearly dependent and it has serious negative effects on the regression model. The regression coefficients may be poorly estimated when multicollinearity exists, which may result in inaccuracy of the regression model. One popular technique to detect multicollinearity is variance inflation factor (VIF). VIF measures how much variance of the estimated regressor coefficient is inflated. Each regressor in the regression model corresponds to one VIF value. This will be very useful to determine whether that regressor is involved in multicollinearity. High VIF indicates the associated regressor may have poor coefficient. VIFs below 3 suggests multicollinearity does not exist [15]. A VIF of one means there is no correlation between the associated regressor and the other regressors.

## IV. DATA-DRIVEN BASED LINEARIZED POWER FLOW MODEL

The regression analysis technique is often used for determining the causal effect relationship between the variables. With power engineering insight, this work uses regression analysis technique to improve the simplified model derived from linearization of the complex full AC power flow model. This section first presents the traditional DC model and the regular linearized AC model; then it proposes the data-driven DC (DDC) model and the data-driven linearized AC model by applying regression analysis to capture the most recent system status in the model.

*A. DC Power Flow Model*

The most common linearized power flow model is the DC model represented by the equation shown below. This DC model applies to the cases when voltage magnitude and reactive power are not of concern. Due to its linearity and low computational burden, the DC power flow model is widely used in academic studies, as well as industrial practice.

$$P_{ij} = \theta_{ij}/x_{ij} \qquad (12)$$

*B. Linearized AC Power Flow Model*

Though DC power flow model has been widely used for dozens of years, it does not apply for the scenarios when voltage and reactive power are of concern. Therefore, this paper first introduces a regular linearized AC power flow model and then proposes a data-driven linearized AC power flow model with better performance and smaller model error.

The nodal voltage magnitude can be written as [16],

$$V_i = 1 + \Delta V_i \tag{13}$$

where $\Delta V_i$ is supposed to be very small since nodal voltage is very close to one in normal situations. Substituting (13) into (1) and (2) and ignoring the second order terms including $\Delta V_i \theta_{ij}$ and $\Delta V_j \theta_{ij}$, we can obtain:

$$P_{ij} = -\theta_{ij} b_{ij} + (\Delta V_i - \Delta V_j) g_{ij} \tag{14}$$
$$Q_{ij} = -(1 + 2\Delta V_i) b_{pi} - \theta_{ij} g_{ij} - (\Delta V_i - \Delta V_j) b_{ij} \tag{15}$$

Then, substituting $\Delta V_i = V_i - 1$, derived from (13) back into (14) and (15), we will obtain the linearized AC power flow equations shown below.

$$P_{ij} = -\theta_{ij} b_{ij} + (V_i - V_j) g_{ij} \tag{16}$$
$$Q_{ij} = b_{pi} - 2V_i b_{pi} - \theta_{ij} g_{ij} - (V_i - V_j) b_{ij} \tag{17}$$

### C. Data-driven Linearized Power Flow Models

In the simplification of the DC model represented by (12), voltage magnitude is assumed to be one per unit all over the entire system, which is one of the main sources why DC power flow model is not very accurate.

Thus, it would be very useful if a better model with similar computational complexity is available. For a given system, the system status including voltage profile may not change significantly, especially in a short term. Thus, this paper presents a data-driven DC power flow model that uses regression analysis technique to capture the system specific condition and reflect it with different coefficients in the model. Thus, as shown in (18), instead of having the proportional parameter being one, we can adjust it and optimize it based on the system most recent status.

$$P_{ij} = K_D \theta_{ij} / x_{ij} \tag{18}$$

Similar to the DDC power flow model, adjustments can be made to the LAC model to incorporate the system most recent status. The data-driven LAC model with coefficients associating to each term is shown below. Model (19) represents regression model $P$ while (20) denotes regression model $Q$. The constant term in (20) is the intercept of the regression model $Q$.

$$P_{ij} = -K_{A1} \theta_{ij} b_{ij} + K_{A2} (V_i - V_j) g_{ij} \tag{19}$$
$$Q_{ij} = b_{pi} - 2K_{A3} V_i b_{pi} - K_{A4} \theta_{ij} g_{ij} - K_{A5} (V_i - V_j) b_{ij} \tag{20}$$

### D. Metrics

The approximately error of the simplified power flow heuristic model is defined in (21) below,

$$\epsilon_1 = \frac{1}{N_1} \sum_{i=1}^{N_1} \left| \frac{x_1^i - x_{AC}^i}{x_{AC}^i} \right| \tag{21}$$

where $x_1$ denotes the state variable for simplified power flow model while $x_{AC}$ denotes the same state variable obtained with the full AC model; $i$ indicates the $i^{th}$ element while $N_1$ denotes the number of elements in $x_1$ or $x_{AC}$.

To quantify how much the proposed data-driven model improves the solution accuracy against the traditional model, a model improvement index, a percentage number like $\epsilon$, is defined in (22) below.

$$\eta_{b,a} = \frac{\epsilon_a - \epsilon_b}{\epsilon_a} \tag{22}$$

In addition to analyzing the average model percentage improvement, the sum and the average of absolute deviations across all buses of interest or all branches of interest is an important indicator on the performance of heuristic models.

The sum of absolute deviation (SAD) is defined in (23) while the average of absolute deviations is defined in (24).

$$s_1 = \sum_{i=1}^{N_1} |x_1^i - x_{AC}^i| \tag{23}$$
$$\gamma_1 = \frac{1}{N_1} \sum_{i=1}^{N_1} |x_1^i - x_{AC}^i| \tag{24}$$

## V. CASE STUDIES

The proposed data-driven DC model and data-driven linearized AC model are tested against the practical TVA system and compared with the traditional full AC power flow model and simplified DC power flow model. This system has 1779 buses and 2301 branches. 72 consecutive hourly cases that covers 3 days' scenarios are used and tested in this work. They are referred to as *hour 1 case*, *hour 2 case*, ..., and *hour 72 case* in this paper.

AC power flow simulation is conducted on *hour 1 case* to obtain the system status that is used as the training data for regression analysis. The other cases are used to validate the proposed models. In this work, R is used as the statistical tool to perform regression analysis [17].

### A. Data-driven DC Model

With the power flow results obtained from the full AC power flow simulation on *hour 1 case*, regression analysis determines the coefficient $K_D$ in (18) to be 1.12, which is above one unit. This is consistent with the fact that the average voltage magnitude for the same *hour 1 case* is 1.04 which is also above one unit.

The coefficient of determination for this regression model is 0.9964. This indicates that the regression model, $P_{ij} = 1.12 \theta_{ij} / x_{ij}$, can explain 99.64% of the variation of the response variable $P_{ij}$. VIF does not apply to regression model with one single regressor. Thus, the DDC model won't have any multicollinearity or overfitting issues as it has only one single regressor.

Fig. 2 shows the plots of different types of residuals against the fitted values of branch active power flows for *hour 1 case*. The residuals in Fig. 2 (a), (b) and (c) correspond to the original residuals, standardized residuals and R-student residuals respectively. Note that the residual scale for Fig. 2 (b) and (c) is about 10 times larger than Fig. 2 (a) since Fig. 2 (b) and (c) show scaled residuals while Fig. 2 (a) does not. In Fig. 2 (a), majority of the original residuals reside within the range of [-0.2, 0.2] and very few of them goes beyond the boundary of [-0.5, 0.5]. This indicates the errors of the regression model are within an acceptance range. Similarly, Fig. 2 (b) and (c) show that the residuals are mostly located within 3 units of the standard deviation. In conclusion, Fig. 2 illustrates the effectiveness of the regressed DDC power flow model.

Note that $K_D \theta_{ij}$ in (18) is mathematically equivalent to $\theta_{ij}$ in (12); in other words, the proposed DDC model will achieve the same $P_{ij}$ while it can improve the solution for $\theta_{ij}$. The power flow solutions obtained from the DC model and the DDC model for the TVA system condition represented by *hour 1 case* are presented in TABLE I. The branch active power $P$ calculated from both models are the same while the improvement on $\theta_{ij}$ is significant; the improvement for branches with flows of over 50 MW is more than 50% on average. As shown Table II, very similar results are observed for *hour 2 case*, which represents a different scenario with the one used to build the DDC model. This further demonstrate that the proposed DDC model can substantially improve $\theta_{ij}$ as compared to the traditional DC model.

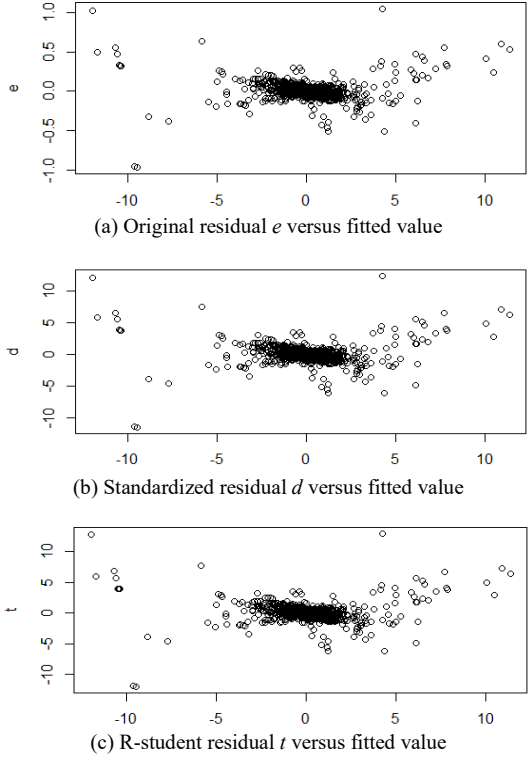

Fig. 2. Plots of residuals against fitted values of branch power for *hour 1 case*

TABLE I
RESULTS WITH DDC AND DC MODELS FOR HOUR 1 CASE

|  |  | Tolerance in MW for $P_{ij}$ | | | |
|---|---|---|---|---|---|
|  |  | 1 | 5 | 10 | 50 |
| # of branches | | 2093 | 1973 | 1746 | 841 |
| $P_{ij}$ | DC | 8.75% | 6.59% | 5.45% | 3.66% |
|  | DDC | 8.75% | 6.59% | 5.45% | 3.66% |
| $\theta_{ij}$ | DC | 22.1% | 17.4% | 13.4% | 11.2% |
|  | DDC | 16.9% | 12.6% | 8.4% | 5.2% |
|  | $\eta_{DLAC,LAC}$ | 23.6% | 27.8% | 37.0% | 53.2% |

TABLE II
RESULTS WITH DDC AND DC MODELS FOR HOUR 2 CASE

|  |  | Tolerance in MW for $P_{ij}$ | | | |
|---|---|---|---|---|---|
|  |  | 1 | 5 | 10 | 50 |
| # of branches | | 2082 | 1954 | 1722 | 815 |
| $P_{ij}$ | DC | 6.96% | 5.00% | 4.28% | 2.94% |
|  | DDC | 6.96% | 5.00% | 4.28% | 2.94% |
| $\theta_{ij}$ | DC | 18.0% | 15.6% | 11.2% | 10.0% |
|  | DDC | 13.8% | 11.6% | 7.18% | 5.20% |
|  | $\eta_{DLAC,LAC}$ | 22.9% | 26.0% | 36.0% | 48.2% |

### B. Data-driven Linearized AC Model

In the DLAC model, two separate regression models are built for branch active power and reactive power respectively. The results of coefficients are presented in Table III. All five coefficients deviate from the value one that is used for the regular LAC model. The 95% confidence interval for each regression coefficient is very narrow, which indicates that the coefficients are very stable and accurate for the given dataset or system conditions.

TABLE III
COEFFICIENTS OF THE REGRESSION MODELS

| Coefficient | | $K_{A1}$ | $K_{A2}$ | $K_{A3}$ | $K_{A4}$ | $K_{A5}$ |
|---|---|---|---|---|---|---|
| Estimate | | 1.12 | 0.98 | 0.87 | 1.02 | 1.01 |
| 95% CI | 2.5% | 1.12 | 0.93 | 0.85 | 1.01 | 1.00 |
|  | 97.5% | 1.12 | 1.03 | 0.88 | 1.02 | 1.04 |

Note that CI denotes confidence interval.

As presented in TABLE IV, the coefficient of determination $R^2$ are 99.74% and 98.71% for the two regression models respectively. This shows that both regression models can explain almost all variance of the branch flows to the means. TABLE IV shows the VIFs for the regressors of both regression models are all very small, well below 3, which means that multicollinearity has little effect on the regression models and there is no overfit issue. TABLE V shows that the residual means of the regression model $Q$ is about five times higher than that of the regression model $P$. This indicates the proposed DLAC model is more accurate for branch active power than branch reactive power.

TABLE IV
RESULTS OF REGRESSION ANALYSIS

| Regression model | $R^2$ | VIF | | | | |
|---|---|---|---|---|---|---|
|  |  | $\theta_{ij}b_{ij}$ | $(V_i - V_j)g_{ij}$ | $V_i b_{pi}$ | $\theta_{ij}g_{ij}$ | $(V_i - V_j)b_{ij}$ |
| P | 0.9974 | 1.05 | 1.05 | NA | NA | NA |
| Q | 0.9871 | NA | NA | 1.06 | 1.20 | 1.16 |

TABLE V
STATISTICAL RESULTS OF MULTIPLE TYPES OF RESIDUALS

| Regression model | e | | d | | t | |
|---|---|---|---|---|---|---|
|  | mean | variance | mean | variance | mean | variance |
| P | 0.0020 | 0.0052 | 0.027 | 1.00 | 0.028 | 1.05 |
| Q | 0.0095 | 0.0024 | 0.19 | 0.96 | 0.19 | 1.01 |

TABLE VI shows the ANOVA analysis for regression model $P$, which are used for testing the significance of regression. As the last column indicates, the possibility of the two regressors being insignificance to the response is negligible. In other words, both the terms $(V_i - V_j)g_{ij}$ and $\theta_{ij}b_{ij}$ are necessary in regression model $P$. This implies that the traditional DC model with only one term $\theta_{ij}b_{ij}$ has obvious room for accuracy improvement. This is consistent with the comparison between TABLE I and TABLE VII. The active power solutions obtained from the LAC and DLAC are very similar, but they are 20% more accurate as compared to the solution obtained with the DC and DDC models. Fig. 3 shows the scatter plot for *hour 1 case*, which indicates the fitted branch active power flows of the proposed DLAC model are very in line with the solutions obtained from the full AC model.

TABLE VII, TABLE VIII and TABLE IX show the statistical results of branch active power, reactive power and bus voltage obtained with DLAC and LAC models on *hour 1 case*

respectively. It is observed that (i) the proposed DLAC model can improve branch reactive power accuracy by 34.5% on branches with flows exceeding 10 MVA and (ii) improve voltage solution by 35.0% against the regular LAC model. As shown in TABLE X, TABLE XI, TABLE XII, very similar observations are made from the results by applying both models to *hour 2 case*. Thus, it is concluded that the proposed data-driven linearized AC model can significantly improve reactive power accuracy by considering system very recent status. The active power derived from the LAC/DLAC model is more accurate than the DC/DDC model.

TABLE VI
ANALYSIS OF VARIANCE FOR RESPONSE VARIABLE $P$

|   | Df | Sum Sq | Mean Sq | F value | Pr (>F) |
|---|---|---|---|---|---|
| $\theta_{ij}b_{ij}$ | 1 | 281.5 | 281.5 | 54167 | < 2.2e-16 |
| $(V_i - V_j)g_{ij}$ | 1 | 4227.1 | 4227.1 | 813356 | < 2.2e-16 |
| Residuals | 2299 | 11.9 | 0.0052 | NA | NA |

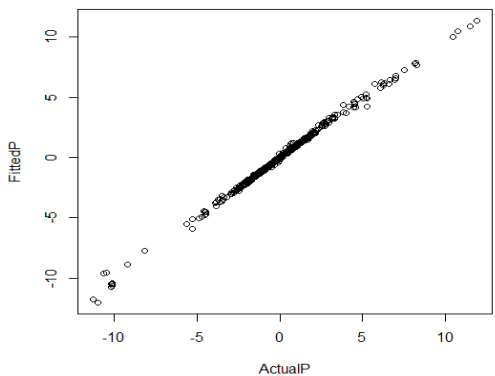

Fig. 3. Branch flows of fitted DLAC model versus AC model for *hour 1 case*

TABLE VII
RESULTS OF ACTIVE POWER WITH DLAC AND LAC FOR *HOUR 1 CASE*

|  |  | Tolerance in MW for $P_{ij}$ | | | |
|---|---|---|---|---|---|
|  |  | 1 | 5 | 10 | 50 |
| # of branches | | 2093 | 1973 | 1746 | 841 |
| $P_{ij}$ | LAC | 6.93% | 5.37% | 4.43% | 3.19% |
|  | DLAC | 7.00% | 5.40% | 4.46% | 3.20% |
| $\theta_{ij}$ | LAC | 18.3% | 14.4% | 12.3% | 11.3% |
|  | DLAC | 12.2% | 8.73% | 6.51% | 4.71% |
|  | $\eta_{DLAC,LAC}$ | 33.6% | 39.2% | 46.9% | 58.5% |

TABLE VIII
RESULTS OF REACTIVE POWER WITH DLAC AND LAC FOR *HOUR 1 CASE*

|  |  | Tolerance in MVAr for $Q_{ij}$ | | | |
|---|---|---|---|---|---|
|  |  | 1 | 5 | 10 | 50 |
| # of branches | | 3860 | 2755 | 2013 | 366 |
| $Q_{ij}$ | LAC | 64.1% | 44.7% | 36.0% | 32.6% |
|  | DLAC | 42.2% | 29.5% | 23.6% | 22.4% |
|  | $\eta_{DLAC,LAC}$ | 34.2% | 34.0% | 34.5% | 31.3% |

It is important to analyze how much improvement the proposed DLAC model has over the conventional LAC model in terms of branch complex power flow that is used to monitor line loading level against line thermal capacity limit. To avoid the effects of small amount of branch flows, the following analysis focuses on branches that have at least 10 MVA. After this filter, the statistics for *hour 1 case* and *hour 2 case* are presented in TABLE XIII. For *hour 1 case*, the average branch flow error is 5.62% for LAC while it is only 4.65% for DLAC, which corresponds to 17.2% model improvement of DLAC over LAC. For *hour 2 case*, the average line flow error is 4.77% for LAC while it is 3.85% for DLAC, which corresponds to 19.1% improvement. The sum of absolute deviation in branch complex power flow (*SADCP*) is also compared for different models. For the traditional LAC model on *hour 1 case*, the *SADCP* is 87.8 MVA over 1876 branches with an average flow error of 4.68 MVA per branch. With DLAC, the *SADCP* drops by 16.0% down to 73.8 MVA, which correspond to 3.93 MVA per line on average. For *hour 2 case*, the *SADCP* is 72.4 MVA over 1864 branches with an average flow error of 3.88 MVA per branch using LAC; With DLAC, the *SADCP* drops by 16.4% down to 60.5 MVA, which correspond to 3.24 MVA per line. Thus, it is concluded that the proposed DLAC model can substantially reduce the branch complex power flow error.

TABLE IX
RESULTS OF VOLTAGE WITH THE DLAC AND LAC MODELS FOR *HOUR 1 CASE*

|  |  | Voltage level / KV | | | | |
|---|---|---|---|---|---|---|
|  |  | All | >=200 | (200, 100] | (100, 20] | < 20 |
| # of buses | | 1877 | 64 | 1205 | 192 | 416 |
| $V_i$ | LAC | 0.0161 | 0.0171 | 0.0188 | 0.0159 | 0.0080 |
|  | DLAC | 0.0105 | 0.0094 | 0.0124 | 0.0106 | 0.0050 |
| $V_i\%$ | LAC | 1.54% | 1.61% | 1.81% | 1.54% | 0.77% |
|  | DLAC | 1.00% | 0.89% | 1.19% | 1.03% | 0.48% |
|  | $\eta_{DLAC,LAC}$ | 35.0% | 44.7% | 34.3% | 33.1% | 37.7% |

TABLE X
RESULTS OF ACTIVE POWER WITH DLAC AND LAC FOR *HOUR 2 CASE*

|  |  | Tolerance for $P_{ij}$ in MW | | | |
|---|---|---|---|---|---|
|  |  | 1 | 5 | 10 | 50 |
| # of branches | | 2082 | 1954 | 1722 | 815 |
| $P_{ij}$ | LAC | 5.08% | 3.98% | 3.50% | 2.53% |
|  | DLAC | 5.10% | 4.01% | 3.51% | 2.53% |
| $\theta_{ij}$ | LAC | 16.4% | 14.0% | 10.8% | 10.2% |
|  | DLAC | 10.4% | 8.42% | 5.55% | 4.40% |
|  | $\eta_{DLAC,LAC}$ | 36.5% | 40.0% | 48.7% | 57.0% |

TABLE XI
RESULTS OF REACTIVE POWER WITH DLAC AND LAC FOR *HOUR 2 CASE*

|  |  | Tolerance for $Q_{ij}$ in MVAr | | | |
|---|---|---|---|---|---|
|  |  | 1 | 5 | 10 | 50 |
| # of branches | | 3811 | 2747 | 1996 | 333 |
| $Q_{ij}$ | LAC | 59.4% | 42.6% | 35.4% | 32.0% |
|  | DLAC | 37.9% | 27.7% | 22.8% | 21.1% |
|  | $\eta_{DLAC,LAC}$ | 36.2% | 35.0% | 35.5% | 34.1% |

Two power flow simulations based on the regular LAC model and the proposed DLAC model are conducted on *hour 1 case*. As compared to the full AC model, the branch complex flow errors in percent for both LAC and DLAC are calculated and presented in Fig. 4 where branches with less than 10 MVA flow and branches with flow errors less than 5% for LAC model are removed in order to clearly show the performance difference between the regular LAC model and the proposed DLAC model. There are 683 branches in Fig. 4 and they are ordered based on flow errors of the proposed DLAC model. It is clearly observed from Fig. 4 that the complex power flow errors of DLAC are well lower than LAC for most branches. To verify the proposed DLAC model on a different system operating condition, similar simulations are performed on *hour 2 case* representing the system scenario of the next following hour and the results are shown in Fig. 5. Very similar observations can be made from Fig. 5, which validates the proposed DLAC model. It is concluded the proposed DLAC model substantially improves the regular LAC model with regression analysis.

TABLE XII
RESULTS OF VOLTAGE WITH DLAC AND LAC FOR *HOUR 2 CASE*

|  |  | Voltage level / KV | | | |
|---|---|---|---|---|---|
|  |  | All | >=200 | (200, 100] | (100, 20] | < 20 |
| # of buses |  | 1877 | 64 | 1205 | 192 | 416 |
| $V_i$ | LAC | 0.0153 | 0.0166 | 0.0180 | 0.0152 | 0.0076 |
|  | DLAC | 0.0097 | 0.0089 | 0.0115 | 0.0099 | 0.0045 |
| $V_i\%$ | LAC | 1.47% | 1.56% | 1.72% | 1.47% | 0.73% |
|  | DLAC | 0.93% | 0.84% | 1.10% | 0.96% | 0.44% |
|  | $\eta_{DLAC,LAC}$ | 36.8% | 46.2% | 36.1% | 34.7% | 39.7% |

TABLE XIII
STATISTICAL RESULTS OF BRANCH COMPLEX POWER WITH FULL AC MODELS WITH A TOLERANCE OF 10 MVA FOR *HOUR 1 CASE* AND *HOUR 2 CASE*

|  | min | max | mean | median | s.d. | # of branches |
|---|---|---|---|---|---|---|
| Hour 1 case | 10.0 | 1189.7 | 89.9 | 49.5 | 135.9 | 1876 |
| Hour 2 case | 10.0 | 1174.9 | 86.9 | 48.0 | 132.3 | 1864 |

With above analysis, it is demonstrated that the proposed data-driven linearized AC power flow model can largely enhance the reactive power profile and voltage profile for the training case on which regression analysis is performed and the case of the very next following hour. Moreover, the proposed DLAC model and the traditional LAC model are tested on 70 more consecutive cases. The results are shown in TABLE XIV and TABLE XV.

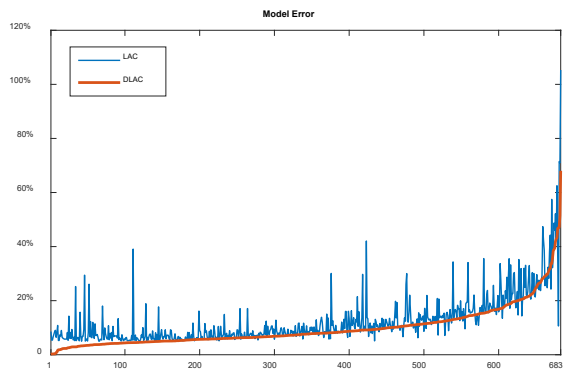

Fig. 4. Branch complex power flow errors of LAC model and DLAC model for *hour 1 case*

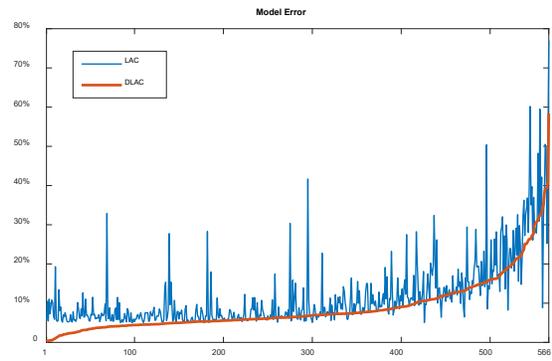

Fig. 5. Branch complex power flow errors of LAC model and DLAC model for *hour 2 case*

TABLE XIV
IMPROVEMENT OF VOLTAGE MAGNITUDE FOR 72 CONSECUTIVE CASES

| hour | LAC | DLAC | $\eta_{DLAC,LAC}$ | hour | LAC | DLAC | $\eta_{DLAC,LAC}$ |
|---|---|---|---|---|---|---|---|
| 1 | 0.0161 | 0.0105 | 35.0% | 37 | 0.0105 | 0.0044 | 58.4% |
| 2 | 0.0153 | 0.0097 | 36.8% | 38 | 0.0110 | 0.0050 | 54.8% |
| 3 | 0.0144 | 0.0085 | 41.2% | 39 | 0.0119 | 0.0060 | 50.1% |
| 4 | 0.0116 | 0.0055 | 52.2% | 40 | 0.0135 | 0.0076 | 43.4% |
| 5 | 0.0095 | 0.0034 | 63.8% | 41 | 0.0145 | 0.0087 | 40.1% |
| 6 | 0.0089 | 0.0029 | 66.9% | 42 | 0.0150 | 0.0095 | 37.1% |
| 7 | 0.0091 | 0.0032 | 65.2% | 43 | 0.0157 | 0.0103 | 34.2% |
| 8 | 0.0090 | 0.0031 | 65.8% | 44 | 0.0167 | 0.0114 | 31.8% |
| 9 | 0.0090 | 0.0029 | 67.3% | 45 | 0.0174 | 0.0122 | 29.9% |
| 10 | 0.0088 | 0.0027 | 68.7% | 46 | 0.0175 | 0.0123 | 29.6% |
| 11 | 0.0091 | 0.0030 | 67.1% | 47 | 0.0165 | 0.0112 | 32.1% |
| 12 | 0.0105 | 0.0043 | 58.9% | 48 | 0.0151 | 0.0097 | 35.6% |
| 13 | 0.0106 | 0.0044 | 58.2% | 49 | 0.0154 | 0.0098 | 36.0% |
| 14 | 0.0117 | 0.0056 | 52.4% | 50 | 0.0145 | 0.0089 | 38.5% |
| 15 | 0.0131 | 0.0070 | 46.3% | 51 | 0.0136 | 0.0076 | 43.8% |
| 16 | 0.0144 | 0.0085 | 41.1% | 52 | 0.0116 | 0.0054 | 53.1% |
| 17 | 0.0159 | 0.0101 | 36.9% | 53 | 0.0101 | 0.0039 | 61.3% |
| 18 | 0.0163 | 0.0108 | 34.1% | 54 | 0.0093 | 0.0032 | 65.5% |
| 19 | 0.0160 | 0.0107 | 33.0% | 55 | 0.0090 | 0.0030 | 66.7% |
| 20 | 0.0185 | 0.0132 | 28.8% | 56 | 0.0088 | 0.0028 | 68.6% |
| 21 | 0.0177 | 0.0125 | 29.4% | 57 | 0.0089 | 0.0027 | 69.8% |
| 22 | 0.0179 | 0.0127 | 29.0% | 58 | 0.0090 | 0.0028 | 68.5% |
| 23 | 0.0173 | 0.0121 | 30.3% | 59 | 0.0090 | 0.0027 | 69.6% |
| 24 | 0.0160 | 0.0107 | 33.1% | 60 | 0.0094 | 0.0031 | 67.1% |
| 25 | 0.0169 | 0.0115 | 31.8% | 61 | 0.0101 | 0.0038 | 62.7% |
| 26 | 0.0158 | 0.0104 | 34.3% | 62 | 0.0108 | 0.0045 | 58.3% |
| 27 | 0.0137 | 0.0079 | 42.0% | 63 | 0.0118 | 0.0055 | 52.9% |
| 28 | 0.0116 | 0.0056 | 51.4% | 64 | 0.0144 | 0.0085 | 41.2% |
| 29 | 0.0102 | 0.0042 | 58.6% | 65 | 0.0138 | 0.0078 | 43.7% |
| 30 | 0.0092 | 0.0034 | 63.6% | 66 | 0.0148 | 0.0089 | 39.8% |
| 31 | 0.0091 | 0.0033 | 63.9% | 67 | 0.0147 | 0.0091 | 38.1% |
| 32 | 0.0090 | 0.0031 | 65.2% | 68 | 0.0150 | 0.0095 | 36.7% |
| 33 | 0.0089 | 0.0029 | 67.4% | 69 | 0.0167 | 0.0112 | 33.1% |
| 34 | 0.0090 | 0.0030 | 66.5% | 70 | 0.0171 | 0.0116 | 32.1% |
| 35 | 0.0092 | 0.0031 | 65.9% | 71 | 0.0165 | 0.0110 | 33.4% |
| 36 | 0.0100 | 0.0039 | 61.1% | 72 | 0.0159 | 0.0102 | 35.8% |

The average voltage improvement with the proposed DLAC model against the regular LAC model over all 72 cases has a mean of 48.7% and a standard deviation of 14.2%; the average branch reactive power improvement with DLAC against LAC is 39.8% with a standard deviation of 4.7%. This demonstrates the proposed DLAC power flow model significantly improve the regular LAC model. The average error of voltage solutions obtained from DLAC is only 0.67% among 72 system scenarios. The standard deviation of voltage error is 0.33%, which indicates the proposed DLAC model is stable and robust over different system conditions. For branch reactive power, the average error is 19.4% with a standard deviation of 6%; though reactive power solution is not as accurate as voltage, the proposed DLAC can provide reactive power information within an acceptable range, which shows its superiority over the traditional DC model.

TABLE XV
IMPROVEMENT OF REACTIVE POWER FOR BRANCHES EXCEEDING 10 MVAR FOR 72 CONSECUTIVE CASES

| hour | LAC | DLAC | $\eta_{DLAC,LAC}$ | hour | LAC | DLAC | $\eta_{DLAC,LAC}$ |
|---|---|---|---|---|---|---|---|
| 1 | 36.0% | 23.6% | 34.5% | 37 | 25.3% | 14.3% | 43.5% |
| 2 | 35.4% | 22.8% | 35.5% | 38 | 26.6% | 15.01% | 43.5% |
| 3 | 33.6% | 20.9% | 37.8% | 39 | 31.5% | 18.7% | 40.8% |
| 4 | 28.8% | 16.8% | 41.9% | 40 | 35.4% | 21.5% | 39.3% |
| 5 | 22.2% | 12.5% | 43.6% | 41 | 36.5% | 22.6% | 38.1% |
| 6 | 22.1% | 12.2% | 45.1% | 42 | 35.9% | 23.3% | 35.1% |
| 7 | 22.5% | 12.3% | 45.4% | 43 | 39.7% | 26.0% | 34.5% |
| 8 | 22.9% | 12.6% | 45.1% | 44 | 39.8% | 26.9% | 32.5% |
| 9 | 23.0% | 12.6% | 45.0% | 45 | 43.7% | 30.1% | 31.1% |
| 10 | 21.9% | 12.4% | 43.6% | 46 | 42.1% | 28.5% | 32.3% |
| 11 | 22.9% | 12.7% | 44.6% | 47 | 38.1% | 25.7% | 32.6% |
| 12 | 27.0% | 14.9% | 44.7% | 48 | 36.9% | 23.5% | 36.5% |
| 13 | 28.2% | 15.8% | 43.8% | 49 | 37.0% | 23.3% | 37.2% |
| 14 | 30.7% | 18.0% | 41.5% | 50 | 33.1% | 20.7% | 37.3% |
| 15 | 35.1% | 21.7% | 38.1% | 51 | 31.3% | 18.8% | 40.1% |
| 16 | 36.1% | 22.5% | 37.8% | 52 | 28.8% | 16.7% | 42.1% |
| 17 | 38.4% | 25.1% | 34.6% | 53 | 25.6% | 14.1% | 44.8% |
| 18 | 40.2% | 26.5% | 34.0% | 54 | 23.3% | 13.0% | 44.2% |
| 19 | 39.7% | 26.6% | 33.0% | 55 | 22.7% | 12.9% | 43.4% |
| 20 | 46.6% | 32.3% | 30.7% | 56 | 21.9% | 12.3% | 44.0% |
| 21 | 44.1% | 30.3% | 31.3% | 57 | 21.2% | 11.7% | 44.8% |
| 22 | 42.0% | 29.1% | 30.9% | 58 | 21.9% | 12.3% | 43.9% |
| 23 | 40.8% | 27.9% | 31.5% | 59 | 21.7% | 12.4% | 42.9% |
| 24 | 39.1% | 26.0% | 33.6% | 60 | 23.1% | 13.0% | 44.0% |
| 25 | 39.6% | 27.0% | 31.9% | 61 | 23.5% | 13.2% | 43.8% |
| 26 | 39.6% | 25.9% | 34.6% | 62 | 25.9% | 14.5% | 44.0% |
| 27 | 32.7% | 19.7% | 39.8% | 63 | 27.3% | 16.3% | 40.2% |
| 28 | 27.0% | 15.8% | 41.6% | 64 | 36.3% | 22.6% | 37.7% |
| 29 | 25.1% | 14.2% | 43.5% | 65 | 34.4% | 20.6% | 40.2% |
| 30 | 23.1% | 12.9% | 44.0% | 66 | 34.4% | 21.6% | 37.2% |
| 31 | 22.7% | 12.9% | 43.1% | 67 | 35.8% | 23.0% | 35.7% |
| 32 | 22.1% | 12.7% | 42.6% | 68 | 37.6% | 23.2% | 38.2% |
| 33 | 21.5% | 12.2% | 43.2% | 69 | 38.9% | 25.9% | 33.2% |
| 34 | 21.7% | 12.5% | 42.3% | 70 | 40.1% | 27.2% | 32.2% |
| 35 | 22.2% | 12.6% | 43.5% | 71 | 38.9% | 25.9% | 33.3% |
| 36 | 23.4% | 13.2% | 43.5% | 72 | 40.5% | 25.8% | 36.3% |

## VI. CONCLUSIONS

Due to nonlinearity and nonconvexity of AC power flow model, the simplified DC model is widely used in both academic and industry. However, DC model does not report any information about reactive power and voltage profile. To maintain linearity while taking reactive power and voltage into account, this work first introduced a regular LAC model that is shown to substantially improve the branch active power over the DC model. Inspired by the fact that system conditions do not change dramatically in a short term, this paper proposes a data-driven linearized AC power flow model by strengthening the regular LAC model with regression analysis.

The results of regression analysis show that the added terms in the branch active power equation for LAC and DLAC against DC model is significant and that there is no overfitting or multicollinearity issue for the DLAC model. Case studies on the large-scale practical TVA system demonstrate the LAC and DLAC models obtain much more accurate branch active power solutions than DC model. The proposed DLAC model can reflect the specific characteristics of a given system and the most recent system condition. Numerical simulations show that the proposed DLAC model substantially improves the reactive power and voltage solutions against the regular LAC model, which demonstrates the performance and effectiveness of the proposed DLAC model.